\documentclass[preprint,showpacs,showkeys,aps,prd,amsfonts,eqsecnum,nofootinbib]{revtex4}
\usepackage{graphicx,epsfig}
\usepackage[dvips]{color}
\usepackage{amsmath, amssymb, graphics}
\newcommand{\vect}[1]{\mbox{\boldmath$#1$}}

\newcommand{\midvect}[1]{\mbox{\boldmath${\textstyle #1}$}}
\def\gsim{\lower0.5ex\hbox{$\:\buildrel >\over\sim\:$}}
\def\lsim{\lower0.5ex\hbox{$\:\buildrel <\over\sim\:$}}
\begin{document}
\preprint{CUMQ/HEP 145}
%
%
\title{\Large  The Casimir Force in Randall Sundrum Models}
\author{Mariana Frank$^a$}
\email[]{mfrank@alcor.concordia.ca}
\affiliation{$^a$Department of Physics, Concordia University, 7141 Sherbrooke St. 
 West, Montreal, Quebec, CANADA H4B 1R6}
\affiliation{$^b$D{\'e}partement de Physique Th{\'e}orique, Universit{\'e} de Gen{\`e}ve, 1211 Gen{\`e}ve 4, Switzerland}
\author{Ismail Turan$^a$}
\email[]{ituran@physics.concordia.ca}
\affiliation{$^a$Department of Physics, Concordia University, 7141 Sherbrooke St.
 West, Montreal, Quebec, CANADA H4B 1R6}
\affiliation{$^b$D{\'e}partement de Physique Th{\'e}orique, Universit{\'e} de Gen{\`e}ve, 1211 Gen{\`e}ve 4, Switzerland}
\author{Lorric Ziegler$^{a,b}$}
\email[]{zieglel0@etu.unige.ch}
\affiliation{$^a$Department of Physics, Concordia University, 7141 Sherbrooke St.
 West, Montreal, Quebec, CANADA H4B 1R6}
\affiliation{$^b$D{\'e}partement de Physique Th{\'e}orique, Universit{\'e} de Gen{\`e}ve, 1211 Gen{\`e}ve 4, Switzerland}

\date{\today}

\begin{abstract}
We discuss and compare the effects of one extra dimension in the Randall Sundrum models on the evaluation of the Casimir force between two parallel plates. We impose the condition that the result reproduce the experimental measurements within the known uncertainties in the force and the plate separation, and get an upper bound $kR\lesssim 20$ if the curvature parameter $k$ of $AdS_5$ is equal to the Planck scale. Although the upper bound decreases as $k$ decreases, $kR\sim 12$, which is the required value for solving the hierarchy problem, is consistent with the Casimir force measurements. For the case where the $5^{th}$ dimension is infinite, the correction to the Casimir force is very small and negligible. 

\end{abstract}
\maketitle
\section{Introduction}\label{sec:intro}

One of the most important questions in particle physics today is how the Standard Model (SM) will be modified at the TeV scale, and how to incorporate gravity into our theories, also known as the problem of the mass hierarchy. The  fundamental problem is why is gravity so weak compared to the other three known fundamental interactions. Gravitational interactions are suppressed by a very high energy scale, with the Planck mass $M_{\rm Pl}\sim 10^{19}$ GeV. In quantum theory, this implies a severe tuning of the fundamental parameters to more than 30 decimal places to keep the values of masses at their observed values.

Early attempts to unify gravity and electromagnetism originated with the Kaluza-Klein (KK) theory \cite{Kaluza:1921tu}, which extended the spacetime to a five dimensional manifold and imposed the condition that the fields should not depend on the extra dimension. With the development of string theory, extra dimensional theories gained a wider acceptance. Unless there are extra space dimensions, string theory is anomalous.

Subsequently, new theories have been developed that explain the mass hierarchy problem by proposing that new dimensions exist, and that the geometry of the extra space dimensions is responsible for this hierarchy. The gravitational field is spread out over the full higher dimensional space, while the SM matter and gauge fields are able to propagate only in a 3 dimensional space, called the 3-brane. Later frameworks suggested that the source of the hierarchy is the strong curvature of the extra dimension. The extra dimensional space can be compactified (made finite)\footnote{There are physically viable scenarios with infinite extra dimensions.}. If the additional dimensions are small enough, the SM particles are allowed to propagate in the extra dimensional manifold (bulk).
 
The first suggestion for generating the hierarchy was to introduce large extra dimensions (LED)  \cite{Arkani-Hamed:1998rs}. If the additional dimensions are too large, this would result in observable deviations from Newtonian gravity, setting a limit on the size of the dimension(s).  In the framework of large extra dimensions, the SM gauge and matter fields are confined to a 3-dimensional brane (3-brane) which exists within a higher dimensional bulk, while only gravity propagates in the extra spatial dimensions which are compactified. The Planck scale $M_{\rm Pl}$ of the 4-dimensional low energy theory is related to the scale where gravity becomes strong in the $4+n$ dimensional space $M_D$ through the volume of the compactified dimensions~$V_n$
\begin{equation}
M_{\rm Pl}^2=V_n M_D^{2+n}\,.
\end{equation}

If $M_D \sim $ TeV, this resolves the hierarchy problem between the Planck and the electroweak scale.
In this model the extra dimensions are flat and of equal size, and the relevant dimension given by the  radius of a toroid, can range in size from 0.1 mm to 10 fermi, and the number of extra dimensions $n$ can vary from 2 to 4. In this scenario $M_{\rm Pl}$ is no longer a fundamental scale, as it is generated by the higher dimensional space.

If the size of the extra dimensions is allowed to be TeV$^{-1}$, the SM fields are also allowed to propagate in the bulk \cite{Antoniadis:1990ew}. The drawback is that, in this scenario, there is no resolution of the hierarchy problem, though it can be incorporated in a framework where the problem is resolved. In models with Universal Extra Dimensions (UED) \cite{Appelquist:2000nn}, all SM fields propagate in the $4+n$ dimensions, with the extra $n$ dimensions taken to be flat and compact, and branes do not need to be present. In the simplest and most popular version, there is only a single extra dimension of size $R$, compactified on an $S_1/Z_2$ orbifold. 

In the Warped Extra Dimensions, the hierarchy between the Planck and the electroweak scales is  generated by introducing compact extra dimensions with large curvatures \cite{Randall:1999vf}. The Randall Sundrum  (RS) model is a 5-dimensional theory compactified on a $S^1/Z_2$ manifold, with bulk and boundary cosmological constants that balance precisely to give rise to a stable 4-dimensional low energy effective theory. The geometry is that of a 5-dimensional Anti-de-Sitter space ($AdS_5$), which is a space of constant negative curvature. 
What prevents gravity from propagating into the extra dimension at low energies is a negative bulk cosmological constant, which is inversely proportional to the radius of curvature.  The background spacetime metric is taken to be
\begin{equation}
ds^2= e^{-2k|y|} g_{\mu \nu} dx^{\mu} dx^{\nu} + \, dy^2\,,
\end{equation}
where the $y$ dependence enters in the so-called ``warp factor" $e^{-2k|y|}$, which is the exponential function of the $5^{th}$ dimension multiplying the usual 4-dimensional Minkowski term. The absolute value of $y$ appears as the extra dimension is compactified on an orbifold. The parameter $k$ governs the degree of curvature of the $AdS_5$ space, assumed to be of the order of the Planck scale. There are two RS models, often denoted as RSI and RSII.

In RSI, there are two branes, localized at $y=0$ and $y=L$, with $Z_2$ symmetry $y\leftrightarrow -y$, $L+y\leftrightarrow L-y$. The 3-branes have equal opposite tensions. The positive tension brane has a fundamental scale $M_{\rm RS}$ and is hidden; SM fields are located on the negative tension brane, which is visible. The exponential warping factor gives rise to an effective scale on the visible brane located at $y=L$
\begin{equation}
M^2_{\rm Pl}=\frac{M_{\rm RS}^3} {k}\left [1-e^{-2kL}\right ]\,.
\end{equation}
A mechanism is needed to recover 4-dimensional general relativity at low-energy, and this corresponds to introducing an extra degree of freedom known as the radion. A suitable choice of $L$ (often taken to be related to the compactified radius, $L=\pi R$) and $k$ allows the KK spectrum to be discrete, and the lowest masses to be of ${\cal O}$(TeV), which predicts different collider signatures at low energies from those of LED. As consistency of the low energy theory requires  $M_{\rm Pl} \sim M_{\rm RS} \sim k$, there are no additional hierarchies in this model. The scale of the 4-dimensional physical phenomena space transverse to the $5^{th}$ dimension is then specified by the warp factor : $\Lambda=M_{\rm Pl}\, e^{-k \pi R}$, and if we take $\Lambda \sim 1$ TeV, then we expect $kR \sim12$.

An interesting alternative of the model occurs if the relevant mass scale parameters ($M_{\rm RS}$, $k$) are taken, instead of being of order $M_{\rm Pl}\, (\sim 10^{18}-10^{19}$ GeV), to be of order TeV \cite{Randall:1999vf,Rattazzi:2003ea}. In this theory the Planck scale is a blue shifted derived scale obtained by rescaling the original action, thus the Lagrangian, by a factor $w=e^{k \pi R}$. Now all the mass parameters are of order TeV, and the Planck mass is generated by 
\begin{equation}
M^2_{\rm Pl}=\frac{(M^{\prime}_{\rm RS})^3} {k^{\prime}}\left [e^{2k^{\prime}L^{\prime}}-1 \right ]\,.
\end{equation}
where $M_{\rm RS}^{\prime}=M_{\rm RS}/w,\,k^{\prime}=k/w$, while $kL=k^{\prime}L^{\prime}$. This scenario is equivalent to ours, since only ratios of mass parameters are observable, and since the two scenarios differ only by a field redefinition \cite{Rattazzi:2003ea}. While we proceed, for the remainder of the paper, to analyze the first scenario where the relevant parameters are taken at the Planck scale, in this latter regime the hierarchy is perhaps more evident and generated in a similar fashion to the LED scenario. For details, see \cite{Rubakov:2001kp}.

In the Randall Sundrum model II (RSII), there is only one positive tension brane. It may be thought of as a limiting case of the previous model, where one brane is located at infinity, $L \rightarrow \infty$. If there is a warped extra dimension, large scales at the Planck brane are shifted at the TeV brane and the relationship between energy scales is given by
\begin{equation}
M^2_{\rm Pl}=\frac{M_{\rm RS}^3} {k}\,.
\end{equation}
Limits on Newton's law set lower bounds on the brane tension and the fundamental scale of RSII \cite{Maartens:2003tw}. The spectrum of RSII is a continuous spectrum of $m>0$ KK modes, and there are no ${\cal O}$(TeV) signatures for this model at the colliders. The infinite dimension makes a finite contribution to the 5-dimensional volume because of the warp factor, and the effective size of the extra dimension probed is $1/k$. 

The possibility of the existence, and the size, of extra dimensions, as well as their geometry influence the structure of the vacuum, in particular the evaluation of the vacuum zero-point energy, known as the Casimir effect  \cite{Casimir:dh}. The research in this area is motivated in two directions. First,  developments in the fundamental area of the structure of the vacuum quantum field theories have been extensively explored, with a view to understand the implications of extra dimensions  \cite{Ponton:2001hq}. Second, several measurements of the attractive force between parallel plates (and other geometries) have firmly established the existence of quantum fluctuations. The level of precision reached by these experiments may be sufficient to test models with different geometries \cite{Bordag}. Previous studies have analyzed cosmological aspects of the vacuum, such as the cosmological constant as a manifestation of the Casimir energy during the primordial cosmic inflation \cite{Peloso:2003nv}. The Casimir energy has been investigated in the context of string theories \cite{Fabinger:2000jd}, and even in Randall Sundrum models, as  means of stabilizing the radion \cite{Garriga:2002vf}. The dynamical Casimir effect has also been discussed in warped braneworlds \cite{Durrer:2007ww}. Recently, the Casimir force for parallel plate geometry has been calculated in UED \cite{Poppenhaeger:2003es,Pascoal:2007uh} and in various other frameworks \cite{Nam:2000cv}.  
  
 To the best of our knowledge, the effects of extra dimensions on the Casimir force and energy have not been calculated in the  warped space time of models RSI and RSII, which is what we propose to do in this present paper. For simplicity, and comparison to experimental measurement, we choose to calculate the Casimir force in RSI and RSII between two parallel plates.

\section{Calculating the Casimir Force in RSI and RSII Models}

In this study, we work in a modified version of the original RS model with at least a massless scalar field living in the bulk. We also adopt the general practice that one could use scalar field analogy for carrying out the calculation of the Casimir force \cite{Milton:2004ya}. Such an approach simplifies the calculation but care must be taken as the actual photon and its boundary conditions could complicate the problem and modify some of the findings here. A bulk scalar field analogy has been considered to evaluate the gravitational Casimir effect in $AdS_5$ brane-world at both zero \cite{Nojiri:2000bz} and finite temperatures \cite{Brevik:2000vt}. The scalar-graviton analogy can in principle be used to calculate the Casimir force due to gravitational field, following a similar procedure as here. 

The KK spectrum for the bulk scalar field has been discussed in the literature, we like to briefly summarize the results here. The equation of motion for a massless bulk scalar field $\Phi$ is
\begin{equation}
g^{\mu\nu}\partial_\mu\partial_\nu\,\Phi + e^{2ky}\partial_y\left(e^{-4ky}\partial_y\Phi\right) =0
\end{equation}
in a 5-dimensional space-time with the background metric $ds^2= e^{-2k|y|} g_{\mu \nu} dx^{\mu} dx^{\nu}-dy^2$. Here $g^{\mu\nu}$ is the usual 4-dimensional flat metric with signature -2. After introducing the KK decomposition by separation of variables, one can consider the zero and non-zero modes separately. If we call the $y$-dependent part of the field as $\chi^{(N)}(y)$, the zero mode solution becomes $\chi^{(0)}(y)=e^{\pm ky}$, adopting modified Neumann boundary conditions \cite{Gherghetta:2006ha} at $y=0,\pi R$. 

The general solution for the non-zero modes can be expressed in terms of Bessel functions of the first and second kind as
\begin{equation}
\chi^{(N\neq 0)}(y)=e^{2ky}\left(a_1 J_2\left(\frac{m_N e^{ky}}{k}\right)+a_2 Y_2 \left(\frac{m_N e^{ky}}{k}\right)\right)\,, 
\label{eq:RSI-mass}
\end{equation}
where $a_1$ and $a_2$ are arbitrary constants and $m_N$ is the effective mass term for the scalar field once we integrate out the $5^{th}$ dimension $y$. 

To satisfy the boundary conditions at both $y=0$ and $y=\pi R$, the argument of the Bessel functions has to satisfy a general equation which reduces to 
\begin{equation}
 \frac{m_N e^{\pi k R}}{k}\approx \pi(N+\frac{1}{4})\,,\;\;\; N\ge 1,
\label{eq:quantmass}
\end{equation}
if we assume $\pi k R\gg1$, which we will throughout this analysis. The approximation in the above equation is valid asymptotically for $N\gg 1$, but is already very accurate even for $N=1$, where the deviation from the actual value is of the order of 3\% \cite{Rattazzi:2003ea}. So, like in the case of Universal Extra Dimensions\footnote{We repeated the Casimir force calculation in Ref.~\cite{Poppenhaeger:2003es} for 5-dimensional UED and agreed with the analytical and numerical results presented there.}, we have a 4-dimensional effective formulation with a discrete KK spectrum for the scalar field. Unlike the UED case, the KK masses here are exponentially suppressed. 

The discussion so far in this section is valid for the RSI since we applied the boundary conditions at both $y=0$ and $y=\pi R$, where the hidden and visible 3-branes are located, respectively. This leads to the quantization in the spectrum. If one considers the RSII model where the brane at $y=\pi R$ is taken to infinity (in this case, the hidden and visible branes are reversed), the solution in Eq.~(\ref{eq:RSI-mass}) will remain the same. However, without the boundary condition at $y=\pi R$, the mass spectrum becomes continuous.

Having summarized the KK spectrum for a bulk scalar field in RSI and RSII models, we continue with the calculation of the Casimir force in RSI first and then later give the result for the force in RSII model. 

The modes of the vacuum in RSI can be expressed
\begin{equation}
 \omega_{nN}=c\sqrt{\vect{k}^{2}_{\perp}+\left(\frac{\pi n}{a}\right)^{2}+m_{N}^{2}}\,,
\end{equation}
where  $m_N$ is given in Eq.~(\ref{eq:quantmass}) and, for later convenience, can be expressed as $\kappa (N+1/4)$ with $\kappa\equiv \pi k e^{-\pi k R}$. We use the Dirichlet boundary conditions on the plates at $z=0,a$. The Casimir energy density per unit plate area will then be given by the usual frequency summation \cite{Pascoal:2007uh}
\begin{equation}\label{mode-sum}
 {\cal E}_{\rm RSI}=2\frac{\hbar}{2}\int\frac{d^{2}\midvect{k}_{\perp}}{(2\pi)^{2}}\left( p \sum^{\infty'}_{n,N=0}\omega_{nN}-\sum^{\infty}_{N=1}\omega_{0N}\right)\;.
\end{equation}
The prime signifies that the term with $n=N=0$ is excluded (for $N=0$, $m_{N}=0$). The parameter $p$ accounts for the polarization of the photon and has to be 3 in 4 space dimensions. The overall factor of 2 is for the volume of the orbifold. We substract the modes polarized in the direction of the brane \cite{Ambjorn:1981xw}. However, even if such a term modifies the energy density, this will not influence the result for the Casimir force since it is independent of the plate separation. Before proceeding further, we carefully rewrite the double sum as 
\begin{eqnarray}
 \nonumber p\sum^{\infty'}_{n,N=0}\omega_{nN}&=&p'\sum^{\infty}_{n=1}\omega_{n0}+p\,c\sum^{\infty}_{n,N=0}\sqrt{\vect{k}^{2}_{\perp}+\left(\frac{\pi n}{a}\right)^{2}+ \kappa^{2}\left(N+ {\frac{  1}{4}}\right)^{2}}\\
 &&-p\,c\sum^{\infty}_{n=0}\sqrt{\vect{k}^{2}_{\perp}+\left(\frac{\pi n}{a}\right)^{2}+\frac{\kappa^{2}}{16}}\;.
\end{eqnarray}
A new polarization factor $p'$ has to be used\footnote{In this case the overall factor of 2 in Eq.~(\ref{mode-sum}) is irrelevant.}, since it corresponds to the usual $4$-dimensional space-time case, i.e. the massless scalar field localized on the brane. We immediately notice that the first term leads to the well known Casimir force per unit area $F_{\rm noRS}=-\hbar c\pi^{2}/(240 a^{4})$, which we do not reproduce here \cite{Casimir:dh}. 

We now proceed separately to the renormalization of the other terms involved. Using the Schwinger representation for the square root and carrying out the integration over $\midvect{k}_{\perp}$,  Eq.~(\ref{mode-sum}) leads to
\begin{eqnarray}\label{epsp}
 \nonumber{\cal E}_{\rm RSI}&=&{\cal E}_{\rm noRS}+\frac{\hbar c\,\pi\Gamma(-3/2)}{\Gamma(-1/2)(2\pi)^{2}}\left( p\sum^{\infty}_{n,N=0}\left[ \left(\frac{\pi n}{a}\right)^{2}+\kappa^{2}\left(N+\frac{1}{4}\right)^{2}\right]^{3/2}\right.\\
 &&\left.-\sum^{\infty}_{N=1}\left[ \kappa^{2}\left(N+\frac{1}{4}\right)^{2}\right]^{3/2}-p\sum^{\infty}_{n=0}\left[ \left(\frac{\pi n}{a}\right)^{2}+\frac{\kappa^{2}}{16}\right]^{3/2} \right)\;.
\end{eqnarray}
The first sum is nothing but a zeta function of Epstein-Hurwitz type
\begin{equation}
 E_{2}(s;a_{1},a_{2};c_{1},c_{2})\equiv\sum_{n_{1},n_{2}=0}^{\infty}\left[ a_{1}(n_{1}+c_{1})^{2}+a_{2}(n_{2}+c_{2})^{2}\right] ^{-s}\;,
\end{equation}
\noindent for which we have the following useful expansion \cite{Elizalde:1992hp}
\begin{eqnarray}
 \nonumber E_{2}(s;a_{1},a_{2};c_{1},c_{2})&=&\frac{a_{2}^{-s}}{\Gamma(s)}\sum_{m=0}^{\infty}\frac{(-1)^{m}\Gamma(s+m)}{m!}\left( \frac{a_{1}}{a_{2}}\right) ^{m}\zeta_{H}(-2m,c_{1})\\
 \nonumber &&\times \zeta_{H}(2s+2m,c_{2})+\frac{a_{2}^{1/2-s}}{2}\sqrt{\frac{\pi}{a_{1}}}\frac{\Gamma(s-1/2)}{\Gamma(s)}\zeta_{H}(2s-1,c_{2})\\
 \nonumber &&+\frac{2\pi^{s}}{\Gamma(s)}\cos(2\pi c_{1})a_{1}^{-s/2-1/4}a_{2}^{-s/2+1/4}\sum_{n_{1}=1}^{\infty}\sum_{n_{2}=0}^{\infty}n_{1}^{s-1/2}\\
 &&\times (n_{2}+c_{2})^{-s+1/2}K_{s-1/2}\left( 2\pi\sqrt{\frac{a_{2}}{a_{1}}}n_{1}(n_{2}+c_{2})\right)\;,
\end{eqnarray}
\noindent where $K_{\nu}$ is the modified Bessel function of the second kind, and $\zeta_{H}(s,q)=\sum_{n=0}^{\infty}(n+q)^{-s}$ is the Hurwitz zeta function.
Our double sum term becomes
\begin{eqnarray}
\!\!\!\!\!\!\!\!\!\! \nonumber E_{2}\left(-\frac{3}{2};\frac{\pi^{2}}{a^{2}},\kappa^{2};0,\frac{1}{4}\right)&=&\kappa^{3}\zeta_{H}(0,0)\zeta_{H}(-3,1/4)+\frac{\kappa^{4}a}{2\sqrt{\pi}}\frac{\Gamma(-2)}{\Gamma(-3/2)}\zeta_{H}(-4,1/4)\\
 &&+\frac{2\kappa^{2}}{a\Gamma(-3/2)\sqrt{\pi}}\sum_{n=1}^{\infty}\sum_{N=0}^{\infty}\frac{(N+1/4)^{2}}{n^{2}}K_{2}(2a\kappa n(N+1/4))\,,
\end{eqnarray}
\noindent where $\zeta_{H}(0,0)=-\frac{1}{2}$.\\
The second sum in Eq. (\ref{epsp}) is simply $\kappa^{3}\zeta_{H}(-3,1/4)$. Finally the last one is again a zeta function of Epstein-Hurwitz type
\begin{equation}
 E_{1}^{c}(s;a_{1};c_{1})\equiv\sum_{n=0}^{\infty}\left[ a_{1}(n_{1}+c_{1})^{2}+c\right] ^{-s}\;,
\end{equation}
\noindent for which we now have the formula \cite{Elizalde:1994gf}
\begin{eqnarray}
\!\!\!\!\!\nonumber E_{1}^{c}(s;a_{1};c_{1})&=&\frac{c^{-s}}{\Gamma(s)}\sum_{m=0}^{\infty}\frac{(-1)^{m}\Gamma(s+m)}{m!}\left( \frac{a_{1}}{c}\right) ^{m}\zeta_{H}(-2m,c_{1})+\frac{c^{1/2-s}}{2}\sqrt{\frac{\pi}{a_{1}}}\frac{\Gamma(s-1/2)}{\Gamma(s)}\\
 &&+\frac{2\pi^{s}}{\Gamma(s)}a_{1}^{-s/2-1/4}c^{-s/2+1/4}\sum_{n_{1}=1}^{\infty}\cos(2\pi n_{1} c_{1})n_{1}^{s-1/2}K_{s-1/2}\left( 2\pi\sqrt{\frac{c}{a_{1}}}n\right),
\end{eqnarray}
\noindent and it gives, for the second term
\begin{equation}
E_{1}^{\kappa^{2}/16}\left(-\frac{3}{2};\frac{\pi^{2}}{a^{2}};0\right)=\frac{\kappa^{3}}{64}\zeta_{H}(0,0)+\frac{\kappa^{4}a}{512\sqrt{\pi}}\frac{\Gamma(-2)}{\Gamma(-3/2)}+\frac{\kappa^{2}}{8a\Gamma(-3/2)\sqrt{\pi}}\sum_{n=1}^{\infty}n^{-2}K_{2}\left( \frac{1}{2}a\kappa n\right)\;.
\end{equation}
We now multiply the energy density ${\cal E}_{\rm RSI}$ between the plates by the area $A$ of one plate to get the total energy
\begin{eqnarray}
\!\!\!\!\!\nonumber E_{\rm RSI}&=&{\cal E}_{\rm RSI}\,A\\
&=&E_{\rm noRS}+\frac{\hbar c\,p\,\pi\kappa^{2}A}{(2\pi)^{2}}\frac{\Gamma(-3/2)}{\Gamma(-1/2)}\left[ \frac{\kappa^{2}a}{2\sqrt{\pi}}\frac{\Gamma(-2)}{\Gamma(-3/2)}\left(\zeta_{H}(-4,1/4)-\frac{1}{256}\right)\right.\nonumber\\
 \nonumber&&+\frac{2}{a\Gamma(-3/2)\sqrt{\pi}}\sum_{n=1}^{\infty}\sum_{N=0}^{\infty}\frac{(N+1/4)^{2}}{n^{2}}K_{2}(2a\kappa n(N+1/4))\\
 &&\left.-\frac{1 }{8a\Gamma(-3/2)\sqrt{\pi}}\sum_{n=1}^{\infty}n^{-2}K_{2}\left(\frac{1}{2}a\kappa n\right)+\frac{\kappa}{128}-\frac{p+2}{2\,p}\,\kappa\,\zeta_{H}(-3,1/4)\right].
\end{eqnarray}
We renormalize the energy by subtracting the one without plates, which compensates exactly the divergent term, i.e. the one containing a $\Gamma(-2)$. To do that, we split the sum into
\begin{equation}
 p\sum^{\infty'}_{N=0}\omega_{0N}=p\,c\sum^{\infty}_{N=0}\sqrt{\vect{k}^{2}+\kappa^{2}(N+1/4)^{2}}-p\,c\sqrt{\vect{k}^{2}+\frac{\kappa^{2}}{16}}\;.
\end{equation}
After analogous steps, we arrive at for the energy density with {\it no plates} ($\rm np$) 
\begin{equation}
\vect{\varepsilon}_{\rm np}=\frac{\hbar c\,p\,\pi^{3/2}\kappa^{4}}{(2\pi)^{3}}\frac{\Gamma(-2)}{\Gamma(-1/2)}\left(\zeta_{H}(-4,1/4)-\frac{1}{256}\right)
\end{equation}
so that the final expression for the total renormalized energy becomes
\begin{eqnarray}
 \nonumber E^{\rm ren}_{\rm RSI}&=&E_{\rm RSI}-\vect{\varepsilon}_{\rm np}\, A\,a \nonumber \\
&=& E_{\rm noRS}+\frac{\hbar c \pi\kappa^{2}A}{(2\pi)^{2}}\frac{\Gamma(-3/2)}{\Gamma(-1/2)}\left[ p\frac{\kappa}{128}-\frac{p+2}{2}\kappa\zeta_{H}(-3,1/4)\right.\nonumber\\
 \nonumber&&+\frac{2 p}{a\Gamma(-3/2)\sqrt{\pi}}\sum_{n=1}^{\infty}\sum_{N=0}^{\infty}\frac{(N+1/4)^{2}}{n^{2}}K_{2}(2a\kappa n(N+1/4))\\
 &&\left.-\frac{p}{a\Gamma(-3/2)8\sqrt{\pi}}\sum_{n=1}^{\infty}n^{-2}K_{2}\left(\frac{1}{2}a\kappa n\right)\right]\;.
\end{eqnarray}
The force per unit plate area is found using the differentiation rule for the Bessel functions $\partial_{z}K_{\nu}(z)=-\frac{1}{2}\left[ K_{\nu-1}(z)+K_{\nu+1}(z)\right]$
\begin{eqnarray}
 \nonumber F_{\rm RSI}&=&-\frac{\partial (E^{\rm ren}_{\rm RSI}/A)}{\partial a}=F_{\rm noRS}-\frac{\hbar c\,p}{4\pi^{2}}\frac{\kappa^{2}}{a^{2}}\left\lbrace \sum_{n=1}^{\infty}\sum_{N=0}^{\infty}\frac{(N+1/4)^{2}}{n^{2}}K_{2}(2a\kappa n(N+1/4))\right.\\
 \nonumber&&+a\kappa\sum_{n=1}^{\infty}\sum_{N=0}^{\infty}\frac{(N+1/4)^{3}}{n}\left[ K_{1}\left(2a\kappa n\left(N+1/4\right)\right)+K_{3}\left(2a\kappa n\left(N+1/4\right)\right)\right]\\
 &&\left.-\frac{1}{16}\sum_{n=1}^{\infty}n^{-2}K_{2}\left(\frac{1}{2}a\kappa n\right)-\frac{a\kappa}{64}\sum_{n=1}^{\infty}n^{-1}\left[ K_{1}\left(\frac{1}{2}a\kappa n\right)+K_{3}\left(\frac{1}{2}a\kappa n\right)\right]\right\rbrace.
\label{eqn:CF-RSI}
\end{eqnarray}
This is our final analytical expression for the Casimir force in the RSI model, to be used for numerical analysis.

After deriving the force in RSI, we proceed with the RSII case. An adaptation to the one brane model, where the position of branes are reversed with respect to RSI, is performed knowing that the spectrum of the KK masses is continuous and consists of all $m>0$. The extra dimensional part of the mode summation in Eq.~(\ref{mode-sum}) is then turned into an integration with measure $dm/k$ \cite{Randall:1999vf}
\begin{equation}
 {\cal E}_{\rm RSII}={\cal E}_{\rm noRS}+\frac{\hbar c p}{2}\int\frac{dm}{k}\int\frac{d^{2}\midvect{k}_{\perp}}{(2\pi)^{2}}\sum_{n=1}^{\infty}\sqrt{\vect{k}_{\perp}^{2}+\frac{\pi^{2}n^{2}}{a^{2}}+m^{2}}\;.
\end{equation}
This corresponds to a Casimir force\footnote{The force in this case can simply be calculated with the use of zeta function renormalization and use of the analytical continuation of the Riemann zeta and Gamma functions.}
\begin{equation}\label{eqn:CF-RSII}
 F_{\rm RSII}=F_{\rm noRS}\left(1+\frac{45 p}{2 \pi^{3}}\zeta(5)\frac{1}{a k}\right)\;.
\end{equation}
where $\zeta(s)$ is the Riemann zeta function.

To complement the analytical study, we perform a numerical analysis of the Casimir force expressions given in Eqs.~(\ref{eqn:CF-RSI}) and (\ref{eqn:CF-RSII}). Let's first consider the RSI case, which is more relevant at low energies. The analytical expression for the force depends on two extra free parameters: in addition to the plate separation $a$, the $AdS_5$ curvature scale $k$, and the size of one extra space dimension $R$ (more precisely $\pi R$). In Eq.~(\ref{eqn:CF-RSI}), the parameter $\kappa \equiv \pi k e^{-\pi k R}$ is used.  

As briefly discussed in the introduction, to solve the hierarchy problem, one needs to assume $kR\sim 12$ if the scale of mass parameters in the hidden brane is assumed to be the Plank mass, $M_{\rm Pl}$. The $AdS_5$ curvature scale $k$ is usually considered at the Planck scale\footnote{There are ways to get $k\sim 1$ TeV if, for example, one formulates the model in some unphysical Galilean coordinates, but the physical conclusions will remain unchanged. See \cite{Kubyshin:2001mc} and references therein for the details.} taken as $10^{19}$ GeV in our numerical analysis. There is no experimental bound for the product $kR$. Our aim is to bound $kR$ from the Casimir force measurements within a 5-dimensional RSI model. For convenience we will use $kR$ and $k$ as the free parameters of the model. 
\begin{figure}[htb]
\begin{center}$
	\begin{array}{cc}
\hspace*{-0.8cm}
	\includegraphics[width=3.3in]{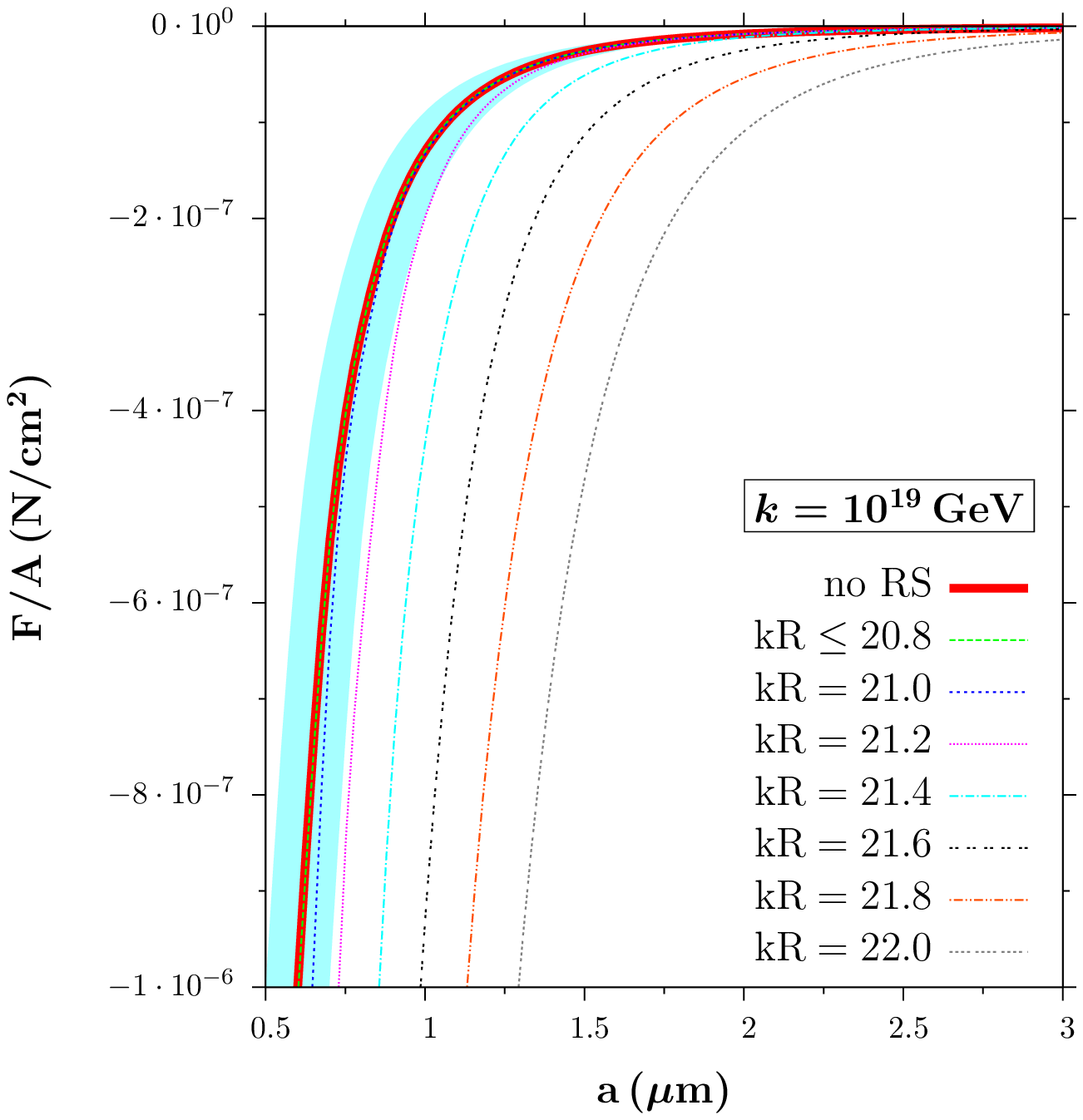} &
	\includegraphics[width=3.3in]{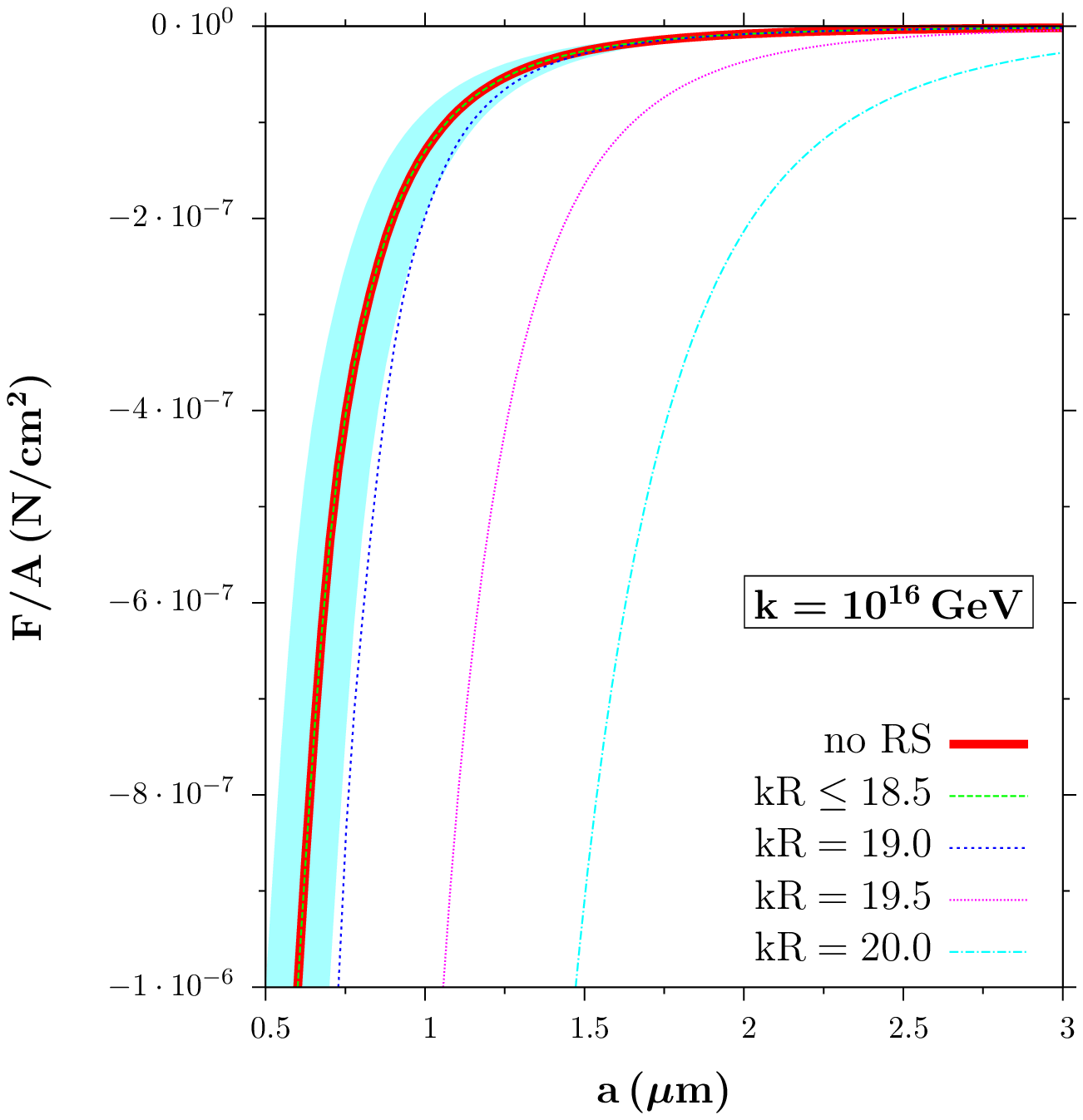} 
	\end{array}$
\end{center}
\vskip -0.2in
      \caption{The Casimir force as a function of the plate separation $a$ in the context of Randall  Sundrum model I for various $kR$ values. In the left panel, the $AdS_5$ curvature scale $k$ is set to $10^{19}$ GeV. In the right panel, it is set to $10^{16}$ GeV. The force for any $kR$ value smaller than 20.8(18.5) in the left(right) panel is represented by a single curve, which coincides with the standard Casimir force curve. The maximum error in $a$ for the standard Casimir force is taken as $\pm 0.1\,\mu m$ and is represented by a color strip.}
\label{fig:RSI-a}
\end{figure}
\begin{figure}[htb]
\begin{center}$
	\begin{array}{cc}
\hspace*{-0.5cm}
	\includegraphics[width=3.0in]{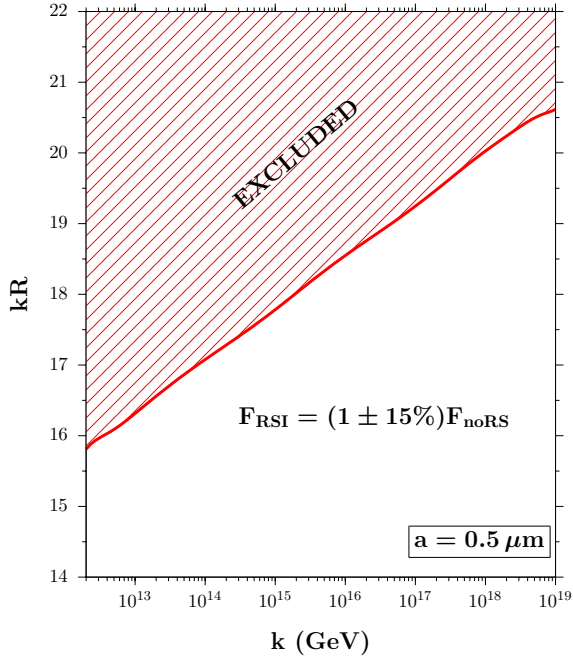} 
	\end{array}$
\end{center}
\vskip -0.2in
      \caption{The contour plot of the Casimir force in RSI in the $k-kR$ plane for a fixed plate separation $a=0.5\,\mu m$. The shaded region is excluded by including a 15\% error in the measured value of the Casimir force.}
\label{fig:RSI-contour}
\end{figure}

In Fig.~\ref{fig:RSI-a}, the Casimir force as a function of the plate separation $a$ is shown for various $kR$ values. In the left panel we set $k$ to the Planck scale, $10^{19}$ GeV. In the right panel, $k$ is taken as $10^{16}$ GeV. Due to the warp factor, the Casimir force is very sensitive to the $kR$ value. For $k=10^{19}$ GeV, any $kR$ value smaller than $20.8$ is represented by one curve which completely coincides with the one without RS contribution, the standard Casimir force. In order to get a rough bound on $kR$ we also include a $\pm 0.1 \mu m$ error in $a$ for the Casimir force measurements. Such a region is shown as a color strip. As seen from the figure, the upper bound for $kR$ is around 21. This value of course depends on the chosen value for $k$. The graph in the right panel is the same as the one on the left but for $k=10^{16}$ GeV. We see that the upper allowed value for $kR$ goes down to 19.  While $kR$ is restricted by our analysis it is seen that $kR\sim 12$ is not disfavored in the one extra dimensional case by the Casimir force measurements. Extension to more than one extra dimension is under investigation \cite{currentwork}.

To further check the sensitivity of the $kR$ upper bound to $k$, we consider the Casimir force in $k-kR$ plane for a fixed plate separation $a=0.5\,\mu m$, presented in Fig.~\ref{fig:RSI-contour}. The region excluded after allowing 15\% uncertainty to the standard Casimir force is shaded. The boundary of the excluded region shows the $k$-dependence of the $kR$ upper bound. Even though $k$ is usually assumed at the Plank scale, we scanned low $k$ values as well, till $k=10^{12}$ GeV. As seen from Fig.~\ref{fig:RSI-contour}, the upper bound of $kR$ can be as low as 16 for $k=10^{12}$ GeV. We should note that such a scenario, if realistic, would require $kR$ values smaller than 12 to solve the hierarchy problem. 

We additionally checked the plate separation  dependence of $kR$ upper bound and found out that it is very weak. For example, if we vary  the plate separation in the $0.5\,\mu m$ to $2\,\mu m$ range, the $kR$ upper bound varies at most by 0.5 and increases as the separation increases.

The situation is quite different for the RSII model. As seen from Eq.~(\ref{eqn:CF-RSII}), the effects of the model on the Casimir force depend on $k$ only and the correction term is ${\cal O}(1/ak)$. Even using the lower limit from the Newton's law on $k$ and for an intermediate plate separation, the correction term will be numerically very small and practically negligible. This is not surprising since it is known that RSII has no low energy effects. 

Finally, we checked numerically the validity of our approximations. We stated that  the expression for the KK modes, $\displaystyle \frac{m_N e^{\pi k R}}{k}\approx \pi(N+\frac{1}{4})$, strictly valid for $N\gg 1$ was in fact valid even for the N=1 state with a 3\% error. We tested to see how much this error would be affected by summing up over all possible modes. The most serious deviation would be obtained for the $N=1$ state, summing over all $n$ modes. In the region where the experimental value of the Casimir force agrees with the Randall-Sundrum prediction, the error is completely negligible. The difference between the force calculated with our approximate formula and the one calculated with the exact formula is largest for $N=1$, $kR=21$ case (for $a=400$ nm) and is at most 3\%. For $N=2$ and $3$ modes, for example, the error drops to 0.3 \% and 0.1\%, respectively.  The final expression for the force displays the feature that, the error introduced by summing over $n$ modes for successive  terms with fixed $N$  becomes less significant than the single largest eigenvalue error. 

\section{Conclusion}
We studied the Casimir force between two parallel plates in the presence of one warped extra dimension of the models proposed by Randall and Sundrum (RS). We analysed first the RSI model, in which two branes (the Planck and TeV branes) exist at a distance $y=\pi R$ from each other. We calculated the Casimir force due to the photon field by using the scalar field analogy, with the fields propagating in the bulk of the warped extra dimension. We introduced the mass of the KK modes and found first the energy, then the force $F$ as functions of $kR$ (with $1/k$ the curvature radius and $R$ the radius of the compactified extra dimension), as well as of $a$, the distance between the plates. We then imposed the condition that the result reproduce the experimental measurements, within the known uncertainties in $F$ and $a$. We found that for a wide range of plausible $k$ values, $kR \lesssim 16$, and most likely  $kR\lesssim 20$ for values of $k$ near the Planck scale. The upper limit on $kR$ decreases when the value of $k$ decreases. The allowed values of $kR$ are consistent with values which support the solution to the hierarchy problem, leading further support to the RSI model. In the limit $R \to 0$ while keeping $k$ fixed we recovered the result for the usual 3-dimensional Casimir force.

We also performed the calculation for the force in the RSII model, in which the 3-brane at $y=\pi R$ is at infinity, and obtained the analytical expression for the force between the plates. The deviation of the 5-dimensional Casimir force from the experimentally measured quantity is unfortunately too small to be probed. Thus this supports previous findings that RSII has no low energy  measurable consequences.

\section{Acknowledgment}
This work is supported in part by NSERC of Canada under the Grant No. SAP01105354. I.T. would like to thank M.~Bleicher for useful communications. L.Z. thanks R.~Durrer for her comments.


\end{document}